\documentclass[sn-mathphys]{sn-jnl-arxiv}

\usepackage{graphicx}
\usepackage{amsmath}
\usepackage{amssymb}
\usepackage{bm}

\jyear{2021}%

\theoremstyle{thmstyleone}%
\theoremstyle{thmstyletwo}%

\theoremstyle{thmstylethree}%

\begin{document}

\title[Deep Spatial and Tonal Optimisation for Diffusion Inpainting]
{Deep Spatial and Tonal Data Optimisation for Homogeneous Diffusion
Inpainting}

\author*{\fnm{Pascal} \sur{Peter}}\email{peter@mia.uni-saarland.de}

\author{\fnm{Karl} \sur{Schrader}}\email{schrader@mia.uni-saarland.de}

\author{\fnm{Tobias} \sur{Alt}}\email{alt@mia.uni-saarland.de}

\author{\fnm{Joachim} \sur{Weickert}}\email{weickert@mia.uni-saarland.de}

\affil{\orgdiv{Mathematical Image Analysis Group, Faculty of Mathematics and 
        Computer Science}, \orgname{Saarland University}, 
        \orgaddress{\street{Campus 
            E1.7},\\ \postcode{66041} \city{Saarbr\"ucken},  \state{Saarland}, 
        \country{Germany}}}


\abstract{Diffusion-based inpainting can reconstruct missing image areas
    with high quality from sparse data, provided that their location and 
    their values are well optimised. This is particularly useful  for
    applications such as image compression, where the original image is known. 
    Selecting the known data constitutes a challenging optimisation problem, 
    that has so far been only investigated with model-based approaches. 
    So far, these methods require a choice between either high quality or high 
    speed since qualitatively convincing algorithms rely on many time-consuming 
    inpaintings. We propose the first neural network architecture that allows 
    fast  optimisation of pixel positions and pixel values for homogeneous 
    diffusion inpainting. During training, we combine two optimisation networks 
    with a neural network-based surrogate solver for diffusion inpainting. This 
    novel concept allows us to perform backpropagation based on inpainting 
    results that approximate the solution of the inpainting equation. Without 
    the need for a single inpainting during test time, our deep optimisation 
    accelerates data selection by more than four orders of magnitude compared 
    to common model-based approaches. This provides real-time performance with 
    high quality results.}


\maketitle


\section{Introduction}\label{sec:intro}
The classical inpainting problem \cite{MM98a, EL99a, BSCB00, GL14} deals with 
input images that have been partially corrupted and aims at reconstructing 
these missing areas. 
However, inpainting can be also useful when the whole image is known. 
For inpainting-based image compression~\cite{Ca88,AG94,DMMH96,GWWB08,WZSG09,
    BHK10,ZD11,GLG12,LSJO12,PHNH16,PKW17,SPME14,BHR21,JS21}, 
the encoder stores only a small percentage of known data from which the
decoder restores the discarded remainder of the image with inpainting.
Some approaches~\cite{Ca88,DMMH96,AG94,BHK10,ZD11,GLG12} such as the pioneering
work of Carlsson~\cite{Ca88} limit the choice of known data to edge locations. 
Following the diffusion-based codec of Gali\'c et al.~\cite{GWWB05,GWWB08}, 
many later approaches \cite{HMWP13,SPME14,PHNH16,PKW17}
rely on careful optimisation of the placement of known data 
in the image domain without the restriction to semantic image features.
Inpainting with partial differential equations (PDEs)~\cite{Ca88} has been able 
to outperform  state-of-the-art codecs: Already simple homogeneous 
diffusion~\cite{Ii62} can compress depth-maps or flow fields better than 
HEVC~\cite{SOHW02} with suitably selected known data  \cite{HMWP13,JPW20,JPW21}.
The problem of choosing the right positions of mask pixels, the so-called 
inpainting mask, is also vital for other applications such as 
denoising~\cite{APW17} or adaptive sampling~\cite{DCPC19}. 
In addition to this \emph{spatial optimisation}, compression also benefits 
from \emph{tonal optimisation}: The values of the known pixels can be adjusted 
to optimise the reconstruction quality as well.

However, even for a simple inpainting operator, spatial and tonal optimisation 
constitute challenging problems. This sparked a plethora of non-neural
approaches \cite{BBBW08,BLPP16,CRP14,CW21,DAW21,DDI06,HSW13,HMWP13,HMHW17,
    HW15,JPW21,KBPW18,MHWT12,MMCB18,Na15,OCBP14,Pe19,PHNH16,PKW17,SPME14}. 
We systematically review those in Section~\ref{sec:related}. Among these 
methods, most require many inpaintings per iteration, which tend to be 
computationally expensive or rely on sophisticated implementations for 
acceleration. 
For instance, probabilistic methods for spatial 
optimisation~\cite{MHWT12,HMHW17} yield high quality masks, but come with a
high computational cost. Theoretical optimality results are rare, but have been 
derived from shape optimisation~\cite{BBBW08} for homogeneous diffusion 
inpainting. This allows near instantaneous spatial optimisation without the 
need for a single inpainting. However, so far, existing discrete 
implementations of this concept do not realise the full potential of the 
theoretical results from the continuous setting.

With our deep data optimisation for homogeneous diffusion inpainting,
we aim for the best of both worlds.
We train neural networks that can optimise both mask positions and 
values without the need for a single inpainting. This allows real-time
performance while maintaining competitive quality for our data selection.
During training, we leverage new hybrid concepts that combine model-based
inpainting with deep learning.


\subsection{Our Contribution}
We propose a deep learning framework for inpainting with homogeneous
diffusion. It is the first neural network approach that allows tonal
optimisation in addition to the selection of spatial positions.
In order to merge model-based inpainting relying on PDEs with 
learning ideas, we propose the concept of a \emph{surrogate solver}: During 
training, a neural network efficiently and accurately emulates diffusion-based 
inpainting while allowing for straightforward backpropagation.
This concept enables us to train spatial optimisation networks that generate
inpainting masks and tonal networks that output optimised known pixel values
for a given mask.

Our contributions extend our previous conference contribution on deep mask 
learning~\cite{APW22} in four distinct ways:
\begin{enumerate}
    \item In addition to a learning approach for spatial optimisation, we also
    propose the first tonal optimisation network.
    \item We improve the network architecture and investigate the impact
    of individual components. 
    \item We discuss and evaluate options to directly generate
    binary masks during training~\cite{Pe22}. 
    With an ablation study we show that our surrogate solver
    is robust under non-binary mask optimisation.
    \item By extending our approach to colour data and performing experiments 
    on significantly larger databases we provide further evidence for the 
    practical relevance of deep data optimisation.
\end{enumerate}
Overall, this constitutes the first comprehensive deep learning framework for 
data optimisation targeted at diffusion-based inpainting. The resulting
mask network marries the quality of probabilistic approaches~\cite{MHWT12} 
with the computational efficiency of instantaneous spatial 
optimisation~\cite{BBBW08}. Similarly, our neural tonal optimisation
consistently offers real-time performance at good quality. Compared
to model-based approaches, its speed is independent of the amount of 
known data in the inpainting mask.
In addition, our networks do not require any parameter tuning
after training, making them attractive for practical applications.


\subsection{Related Work}
\label{sec:related}
In the following, we discuss prior work for spatial and tonal 
optimisation, as well as related deep learning approaches.

\subsubsection{Spatial Optimisation}

Finding good positions for sparse known pixels constitutes a challenging 
optimisation problem that has sparked significant research activities. 
In the following, we mostly focus on approaches for diffusion-based inpainting,
but there are also more broadly related works, for instance the free
knot problem for spline interpolation. For instance,
Sch\"utze and Schwetlick~\cite{SS03} have proposed a data selection 
algorithm for the 2-D setting which can also be 
applied to images.
Model-based methods for diffusion inpainting 
can be organised in four categories.

\begin{enumerate}
    \item \emph{Analytic Approaches.} From the theory of shape optimisation,    
    Belhachmi et al.~\cite{BBBW08} derived optimality statements in the 
    continuous setting. In practice, these can be approximated by dithering
    the Laplacian magnitude of the input image. This approach does not require 
    inpainting to find the mask pixels and is therefore very fast. However, the 
    dithering yields only an imperfect approximation with limited 
    quality~\cite{HMHW17,MHWT12}.
    
    \item \emph{Nonsmooth Optimisation Strategies.} Combining concepts from
    optimal control with sophisticated strategies such as primal-dual solvers,
    multiple works~\cite{BLPP16,CRP14,Na15,HSW13,OCBP14} leverage nonsmooth 
    optimisation for the selection of mask positions. These produce 
    results of high quality, but are difficult to adapt to different 
    inpainting operators. Moreover, they do not allow to specify the target 
    amount of mask points a priori. For applications in compression, the 
    non-binary masks need to be binarised, which reduces quality and requires 
    tonal optimisation \cite{HW15} for good results.
    
    \item \emph{Sparsification Methods.} Mainberger et al.~\cite{MHWT12} have 
    proposed \emph{probabilistic sparsification (PS)} to tackle the 
    combinatorial complexity of spatial optimisation. They start with a full 
    mask and iteratively remove candidate pixels. Among those candidates
    the algorithm discards a fraction of pixels with the smallest 
    inpainting error permanently, while returning the remainder to the mask. 
    This process is repeated until the desired percentage of mask points,
    the target density, is achieved. Besides good quality, this approach can
    easily be adapted to many different inpainting operators, including 
    inpainting 
    with 
    PDEs~\cite{MHWT12,HMHW17} or interpolation on 
    triangulations~\cite{DDI06}. This flexibility and quality comes at the cost 
    of many inpainting operations. Nonetheless, sparsification is popular and 
    widely used due to its advantages and its simplicity.
    
    \item \emph{Densification Approaches.} For applications such as compression 
    or denoising, low densities are required. In such cases it can make sense 
    to start with an empty mask and fill it successively instead of using 
    sparsification. Such strategies~\cite{CW21,DAW21,KBPW18} share the benefits 
    of simplicity, good quality, and broad applicability with sparsification. 
    They have been successfully used for diffusion-based~\cite{CW21,DAW21} and 
    exemplar-based~\cite{KBPW18} inpainting operators. For compression, 
    densification also has been applied to data structures such as subdivision 
    trees instead of individual pixels \cite{DNV97,GWWB08,PHNH16,SPME14}. 
    However, 
    all of these strategies still require a significant amount of inpaintings 
    to 	obtain masks of sufficient quality.
\end{enumerate}

The approaches of Categories 3 and 4 are greedy strategies that 
can become stuck in local minima. 
To address this problem, a relocation strategy, the so-called 
\emph{nonlocal pixel exchange} (NLPE)~\cite{MHWT12} has been proposed as a 
post-processing. It is a probabilistic method that iteratively swaps point 
locations randomly with heuristics for candidate selection based on the 
inpainting error. While it can yield significant additional improvements, it 
also requires more inpaintings and tends to converge slowly. Similar strategies
have also been used for interpolation on triangulations \cite{MMCB18}.

Note that the approach from Category 1 is the only one to require no 
inpaintings or complex solvers. Unfortunately, this near instantaneous spatial 
optimisation yields clearly worse results in terms of quality than the methods
from Categories 2--4. With our deep learning framework, we aim at achieving the 
best of both worlds: Fast spatial optimisation without the need for any 
inpaintings while producing results of a quality comparable to Categories 2--4.

\subsubsection{Tonal Optimisation}

So far, we have discussed methods that focus on finding optimal 
positions at which the original image data is kept. However, in a data 
optimisation scenario, we are not confined to selecting the location, but
can also alter the value of mask pixels. This tonal optimisation introduces 
errors at mask pixels  if those lead to a more accurate reconstruction 
in larger missing areas. Also for tonal optimisation, one can distinguish
several categories:

\begin{enumerate}
    \item \emph{Least Squares Approaches.} 
    For spatially fixed mask pixels, tonal optimisation leads to a least 
    squares problem. The resulting linear system of equations is given
    by the normal equations. It has as many unknowns as mask pixels.
    The system matrix is a quadratic, dense matrix that is symmetric
    and positive definite~\cite{MHWT12}. 
    
    To solve it numerically, various algorithms can be applied. Direct methods 
    include Cholesky, LU, and QR  factorisations, while conjugate gradients and 
    the LSQR algorithm constitute suitable iterative approaches \cite{Bj96a}. 
    Other iterative methods that have been used for tonal optimisation are the 
    L-BFGS algorithm \cite{CRP14} and a gradient descent with cyclically 
    varying step sizes \cite{HMHW17}. All of these approaches suffer from
    the fact that they require to store the full matrix, which can become
    prohibitive for masks with too many pixels. 
    
    A potential remedy of this memory restriction consists of subsequently 
    computing a so-called inpainting echo in a mask pixel \cite{MHWT12}. 
    It describes the influence of the mask pixel on the final inpainting 
    result and can be used to adjust the grey or colour value accordingly. 
    Doing this in random order for all mask pixels can be interpreted as a 
    randomised Gauss--Seidel or SOR iteration step. If one does not store all 
    inpainting echoes but computes them again in each iteration step, one 
    achieves 
    low memory requirements at the expense of a long runtime.
    
    Discrete Green's functions offer another way to decompose
    the inpainting problem into pixel-wise contributions~\cite{HPW15}. 
    From this dictionary, the inpainting result can be assembled with simple
    linear superposition. Hoffmann~\cite{Ho17} have used this property 
    to derive an alternative least squares formulation for tonal optimisation 
    which can be solved efficiently with a Cholesky solver. While its solution
    is equivalent to the direct least squares approach, it benefits
    from speed-ups for low amounts of mask pixels which are represented
    by only a few entries from the Green's function dictionary. 
    
    A recent alternative goes back to Chizhov and Weickert~\cite{CW21}. It uses
    nested conjugate gradient approaches within a finite element framework
    and it is both efficient w.r.t. memory and runtime.
    
    \item \emph{Nonsmooth Optimisation Methods.} Hoeltgen and Weickert 
    \cite{HW15} have shown that thresholded non-binary spatial mask 
    optimisation 
    \cite{BLPP16,CRP14,Na15,HMHW17,OCBP14} is equivalent to a combined 
    selection of 
    binary masks and a tonal optimisation. Thus, the previously discussed 
    nonsmooth strategies also indirectly perform tonal optimisation. However,
    this is inherently coupled to a spatial optimisation with the advantages and
    drawbacks described in the previous section.
    
    \item \emph{Localisation Approaches.} Since the influence of a single mask 
    pixel mainly affects its local neighbourhood, tonal optimisation can be 
    sped 
    up by localisation. Strategies exist for localised operators such 
    as Shepard interpolation with truncated Gaussians~\cite{Pe19}, 1-D linear 
    interpolation \cite{PCW19}, or smoothed particle hydrodynamics 
    \cite{DAW21}. Other approaches limit the influence artificially by 
    subdivision trees \cite{PKW17} or segmentation \cite{HMWP13, JPW21}.
    
    \item \emph{Quantisation-based Strategies.} All compression codecs
    rely on quantisation, the coarse discretisation of the colour domain.
    It can be beneficial to directly take quantisation into account during 
    tonal optimisation instead of applying it in postprocessing.
    Thereby, one replaces the continuous optimisation problem by a 
    discrete one. To this end, Schmaltz et al.~\cite{SPME14} proposed a simple 
    strategy that visits pixels in random order and changes their values if 
    increasing or decreasing the quantisation level yields a better results. 
    Peter 
    et al.~\cite{PHNH16} instead augment the Gauss-Seidel strategy with echoes 
    \cite{MHWT12} with a projection to the quantised grey levels. For 
    interpolation 
    on triangulations, Marwood et al.~\cite{MMCB18} use a stochastic approach 
    that 
    randomly assigns different quantisation levels in combination with spatial 
    optimisation.
\end{enumerate}

In addition to tonal optimisation itself, there are also related strategies.
Gali\'c et al.~\cite{GWWB08} proposed an early predecessor that modified tonal 
values to avoid singularities in PDE-based inpainting. To avoid visually
unpleasant singularities at mask pixels, Schmaltz et al.~\cite{SPME14} use 
interpolation swapping: After the initial inpainting, they remove disks around 
the known data and use the more reliable reconstruction for a second inpainting.

The tonal category 1 is restricted to linear diffusion operators, including 
homogeneous diffusion. Category 2 marks the indirect tonal optimisation 
performed by nonsmooth spatial methods and categories 3 and 4 are mainly 
relevant for practical applications in compression. We aim at providing a 
neural network alternative to Category 1 methods for homogeneous diffusion 
inpainting. As for spatial inpainting, our goal is to propose a deep 
optimisation approach that offers high speed at good quality.

\subsubsection{Relations to Deep Learning Approaches}

To our best knowledge, deep learning approaches for sparse data optimisation 
are still very rare and so far, only spatial optimisation has been covered at 
all. Dai et al.~\cite{DCPC19} have proposed a deep learning method for
adaptive sampling that trains an inpainting and an optimisation network 
separately. Joint training for spatial optimisation and inpainting with 
Wasserstein GANs was introduced by Peter~\cite{Pe22}. Both approaches differ 
significantly from the current one, since they aim at learning both a spatial 
optimisation CNN and the inpainting operator. In contrast, we optimise known 
data for model-based diffusion inpainting with a surrogate solver for 
homogeneous diffusion inpainting. Moreover, our deep data selection is the 
first to consider both spatial and tonal optimisation.

In addition, a plethora of deep inpainting methods exist (e.g. 
\cite{LJXY19,PKDD16,XXC12,YLLS17,YLYS18,WZNL+21,WZZ21}). A full review is 
beyond the scope of this paper, because these approaches do not consider any 
form of data optimisation. Since the selection of known data is decisive
for the quality of inpainting-based compression, the current lack of research 
in this direction is the primary reason why deep inpainting has not played a 
role in this area, yet.

\subsection{Organisation of the Paper}
After a brief review of diffusion-inpainting and model-based optimisation in 
Section~\ref{sec:review}, we propose our deep mask learning approach in 
Section~\ref{sec:ours}. Section~\ref{sec:experiments} provides an ablation 
study and a comparison to model-based data optimisation. We conclude our paper 
with a summary and outlook on future work in Section~\ref{sec:conclusion}.



\begin{figure}[t]
    \tabcolsep1pt
    \centering
     \begin{tabular}{cccc}
        \includegraphics[width=0.23\linewidth]{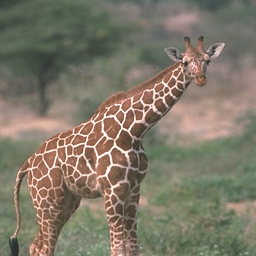}
        &
        \includegraphics[width=0.23\linewidth]{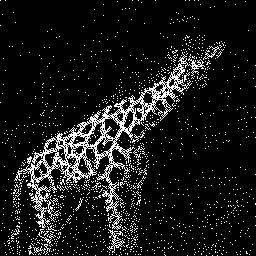}
        &
        \includegraphics[width=0.23\linewidth]
        {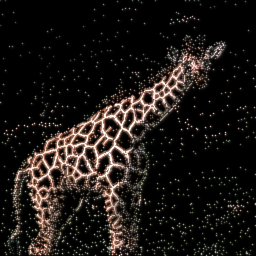} &
        \includegraphics[width=0.23\linewidth]{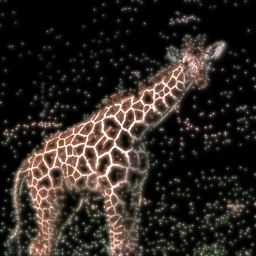}   
        \\[-1mm]    
        original & mask & $t=0$ & $t=1$ \\[1mm]
        \includegraphics[width=0.23\linewidth]{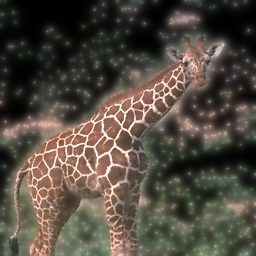}
        &
        \includegraphics[width=0.23\linewidth]{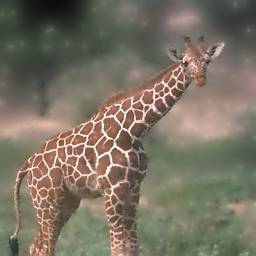}
        &
        \includegraphics[width=0.23\linewidth]{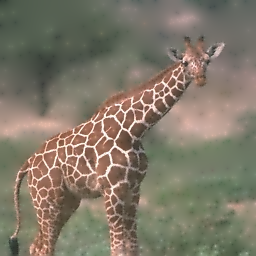} &
        \includegraphics[width=0.23\linewidth]{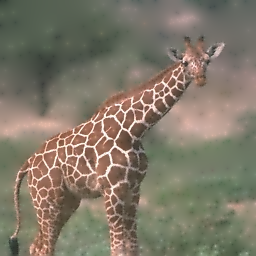}  
        \\[-1mm]  
        $t=10$ & $t=100$ & $t=1000$ & $t=10000$ \\
    \end{tabular}
\\[1mm]
    \caption{\textbf{Image Evolution of Homogeneous Diffusion Inpainting.}
    This 
    figure shows a reconstruction of image 130014 from the BSDS500 
    database~\cite{AMFM11} cropped to size $256 \times 256$.  White mask pixels 
    indicate a total of $10\%$ known data. At 
    time $t=0$, we assign the original pixel values to known areas and 
    initialise the unknown regions with zero (black). 
    For $t\rightarrow\infty$, diffusion 
    propagates the known values and yields the inpainted image as the steady 
    state.
    \label{fig:evolution}}
\end{figure}

\section{Diffusion-based Inpainting and Data Optimisation}\label{sec:review}

Consider a grey value image $f:\Omega\rightarrow \mathbb R$ that is only known 
on the inpainting mask, a subset $K \subset \Omega$ of the rectangular image 
domain $\Omega \subset  \mathbb  R^2$. Diffusion-based inpainting 
\cite{Ca88,WW06} 
reconstructs the missing areas $\Omega \setminus K$ by propagating the
information of the fixed known pixels from $K$ over the diffusion time $t$. 
The inpainted image is the steady state  $t \rightarrow \infty$ of this 
evolution. Fig.~\ref{fig:evolution} illustrates such a propagation over 
time. 
For our inpainting purposes, we are only interested in the steady state 
and not the intermediate steps of the evolution.
 
There are sophisticated anisotropic diffusion 
approaches~\cite{WW06,GWWB08,SPME14,PKW17,JS21b} 
that adapt the amount of propagation in different directions to the image 
structure and can achieve results of very good quality even if the dataset $K$ 
is not highly optimised. However, in the following, we consider  simple  
homogeneous diffusion~\cite{Ii62} for inpainting. It is parameter-free and can
achieve surprisingly high quality for a well-optimised dataset.
In this case, the inpainted image  $u$ fulfils the inpainting equation
\begin{equation}\label{eq:inp}
    \left(1 - c\right) \Delta u - c \left(u - f\right) = 0 \, ,
\end{equation}
which arises as the steady state if one inpaints with the
homogeneous diffusion equation $\partial_t u = \Delta u$. Here,
$\Delta u = \partial_{xx} u + \partial_{yy} u$ denotes the Laplacian
and $c$ is a binary confidence function with $c(\bm x)=1$ for known data in $K$ 
and  $c(\bm x) =0$ otherwise. At the image boundaries $\partial \Omega$ we 
impose reflecting boundary conditions. Note that it is also possible to use 
non-binary confidence values \cite{HW15}, which we will do in 
Section~\ref{sec:nonbinary}. Since homogeneous diffusion is a linear operator, 
colour inpainting is implemented by channel-wise processing.

In practice, we implement this method on a discrete input image $\bm f \in 
\mathbb{R}^{n_x n_y}$ with resolution $n_x \times n_y$. Discretising 
Eq.~\ref{eq:inp} with finite differences leads to a linear system of equations.
Then, reconstructing the image $\bm u \in \mathbb{R}^{n_x n_y}$ is achieved 
with a suitable numerical solver. 

The discrete problem of mask optimisation for homogeneous diffusion inpainting 
consists in finding the binary mask $\bm c \in \{0,1\}^{n_x n_y}$ with a 
user-specified target density $d$ such that $\|\bm c\|_1/(n_x n_y) = d$
where $\|\cdot\|_1$ denotes the 1-norm. This density can be seen as a budget 
that specifies the percentage of image pixels that should be contained in the 
final mask.

For comparisons, we consider the \emph{analytic approach} of Belhachmi et 
al.~\cite{BBBW08}. It is based on results from the theory of shape optimisation 
that demonstrate that mask pixels should be placed at locations of large 
Laplace magnitude. In the discrete setting, they use a Floyd-Steinberg 
dithering \cite{FS76} of the Laplace magnitude. This leads to an imperfect, but 
very fast approximation of the theoretical optimum. This algorithm is a 
representative for simple approaches that do not require any inpaintings to 
determine the optimised mask. 

As a prototype for better performing mask optimisation algorithms, we consider 
the widely used \emph{probabilistic sparsification} of Mainberger et 
al.~\cite{MHWT12}.  It yields better results than the analytic approach by 
taking the discrete nature into account directly and greedily removing pixels 
that are not important for the reconstruction. It starts with a full inpainting 
mask. In each iteration, it removes a fraction $p$ of candidate pixels from the 
mask. After an inpainting with the new mask, it analyses the local inpainting 
error: Candidate pixels which have a high local inpainting error are hard to 
reconstruct and should thus not be removed. Therefore, the algorithm adds back 
the fraction $q$ of candidates with the largest errors. The iterations are 
repeated until the target density $d$ is reached.

\begin{figure}[t]
    \tabcolsep1pt
    \centering
    \begin{tabular}{cccc}
        random & AA~\cite{BBBW08} & PS~\cite{MHWT12} &  
        PS+NLPE~\cite{MHWT12} \\
        \includegraphics[width=0.23\linewidth]{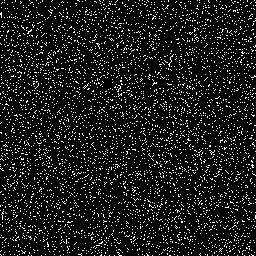}
        &
        \includegraphics[width=0.23\linewidth]{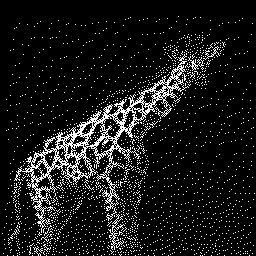}
        &
        \includegraphics[width=0.23\linewidth]{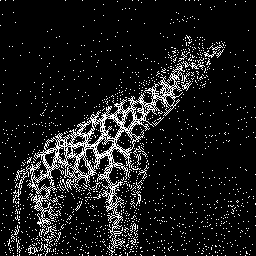} &
        \includegraphics[width=0.23\linewidth]{images/giraffe_nlpe_mask.png}   
        \\    
        \includegraphics[width=0.23\linewidth]{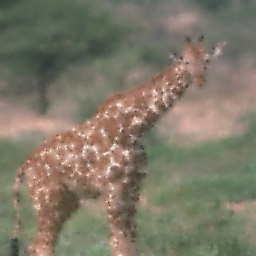}
        &
        \includegraphics[width=0.23\linewidth]{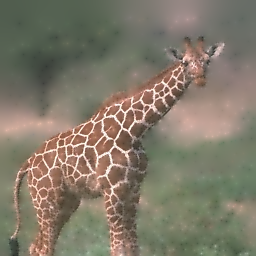}
        &
        \includegraphics[width=0.23\linewidth]{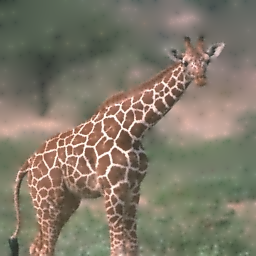} &
        \includegraphics[width=0.23\linewidth]{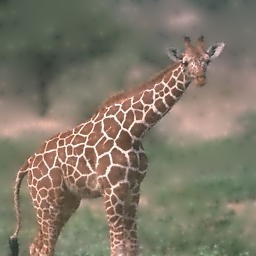}  
        \\
        PSNR: $23.06$ & PSNR: $28.57$ & PSNR: $31.24$ &PSNR: $32.69$ \\
    \end{tabular}\\[1mm]
    \caption{\textbf{Spatial Optimisation Techniques.} For 
    reference, we consider a uniformly random mask with $10\%$ known data and 
    the  corresponding reconstruction of image 130014 from 
    Fig.~\ref{fig:evolution} with homogeneous diffusion inpainting.
     The analytic approach~\cite{BBBW08} (AA) already yields a significant 
    improvement over the random mask. Probabilistic sparsification (PS) and 
    non-local pixel exchange (NLPE) \cite{MHWT12} refine the background as well 
    as the fur patterns of the giraffe. A gain of more than 9dB PSNR 
    illustrates the vital importance of spatial optimisation for homogeneous 
    diffusion inpainting.\label{fig:spatial_model}
}
\end{figure}

Further improvements can be achieved with the \emph{nonlocal pixel 
exchange}~\cite{MHWT12}. 
It is designed to escape from potential local minima by moving a set 
of $p$ candidate locations from the inpainting mask to locations in the
unknown image areas. If this positional exchange improves the overall 
inpainting, it is maintained, otherwise it is reverted. While this guarantees 
that mask quality cannot deteriorate, each step requires an inpainting and 
therefore, convergence tends to be slow.

In Fig.~\ref{fig:spatial_model}, a comparison of the three aforementioned 
spatial optimisation techniques 
 with a uniformly random mask highlights their 
significant impact. Carefully optimised known data are integral for good 
 inpainting results.

Since we consider homogeneous diffusion and do not require quantisation, we use 
a least squares approach for tonal optimisation. Due to the
similar quality of the tonal methods from Section~\ref{sec:related}, we choose 
the Green's formulation by Hoffmann et al.~\cite{Ho17} equipped with a Cholesky 
solver. It offers good quality at fairly low computational cost, 
in particular for very sparse masks.

In the following sections we introduce a deep learning approach that does not 
require inpaintings during spatial or tonal optimisation and approximates the 
quality of probabilistic methods and model-based tonal optimisation.


\begin{figure*}[t]
    \centering
    \includegraphics[width=\linewidth]{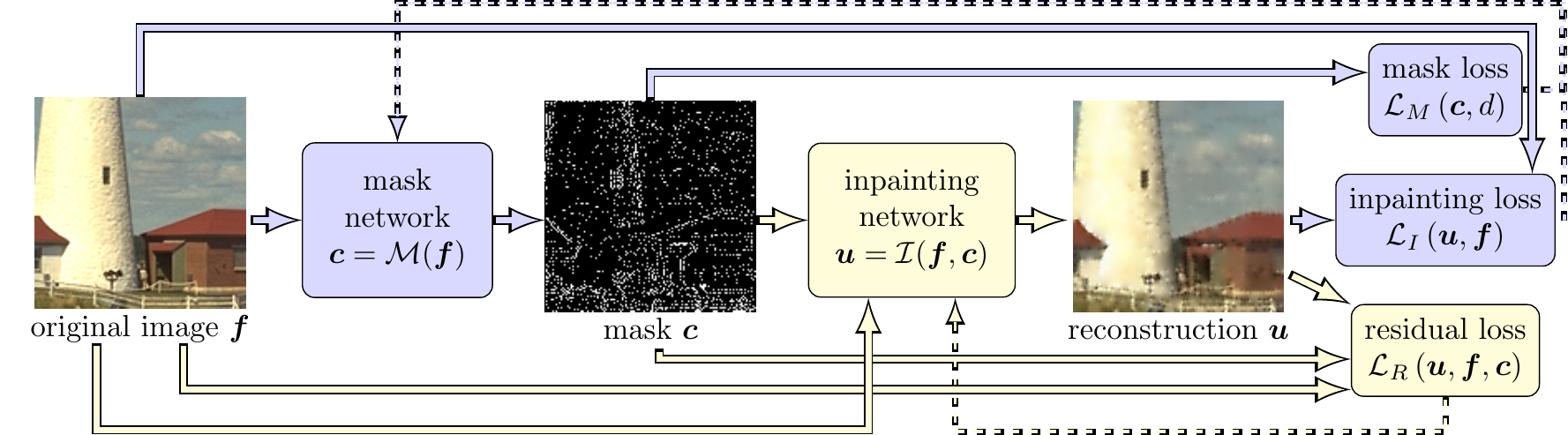}\\
    (a) training of the mask network \\[2mm] 
    \includegraphics[width=\linewidth]{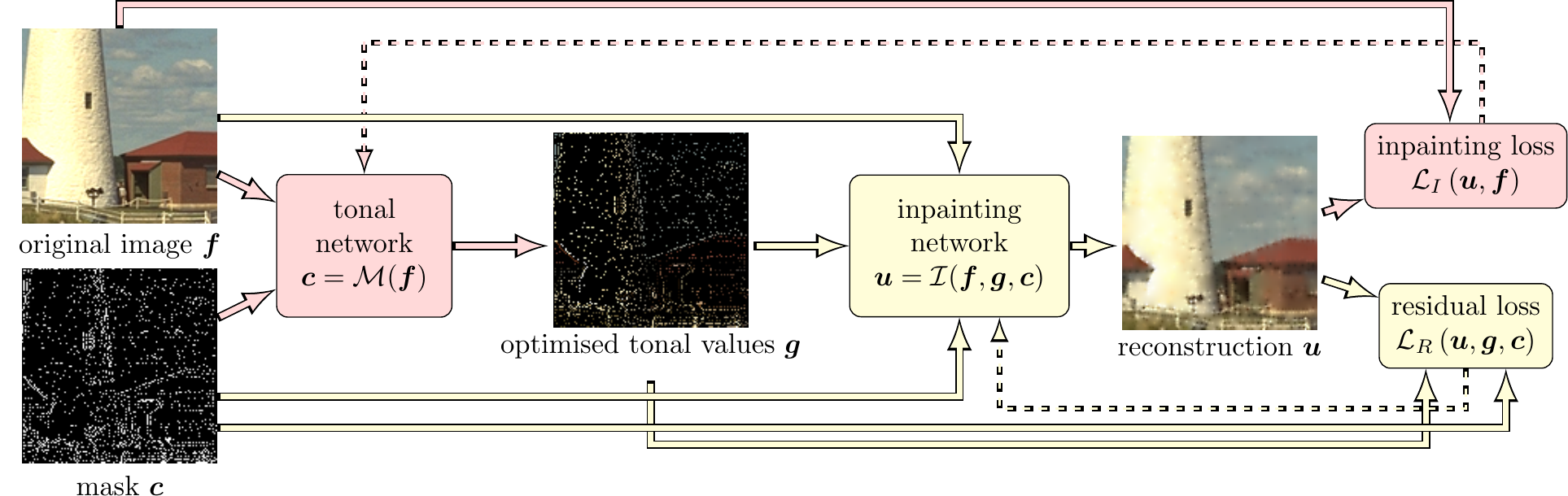}\\[-2mm]
    (b) training  of the tonal network \\[1mm]
    \caption{\textbf{Structure of Our Deep Tonal Optimisation Framework}.
        The surrogate inpainting network and its associated residual loss are 
        marked in yellow, the mask network and loss in blue, and the tonal 
        network 
        in red. Forward passes between the networks are indicated by 
        solid arrows, while dashed arrows represent backpropagation.
        \label{fig:model_structure}}
\end{figure*}


\section{Spatial and Tonal Optimisation with Surrogate 
    Inpainting}\label{sec:ours}

In this section, we describe the three types of networks that act as 
the building blocks for our neural data optimisation framework. The centrepiece 
required for our different pipelines is the \emph{surrogate inpainting 
    network}. It approximates inpainting with homogeneous diffusion by 
    minimising
the residual of the inpainting equation. We only use it during training. 
Its sole purpose is to act as a fast approximate solver for the inpainting
problem that is still differentiable and allows backpropagation.

For the data optimisation, we consider a \emph{mask network} for spatial 
optimisation and a \emph{tonal network} for optimisation of the pixel values.
Each of them is trained together with a separate surrogate inpainting network.
Both data optimisation networks minimise the inpainting error w.r.t. the 
reconstruction by the respective surrogate solver. 

In addition, the mask network requires a separate loss to approximate the 
intended mask density $d$. The macro architecture of our spatial approach with 
can be found in Fig.~\ref{fig:model_structure}(a).

For the tonal setting in Fig.~\ref{fig:model_structure}(b), we have a similar
overall setup. However, here the binary masks are already part of the training 
dataset. In practice, we use our tonal network to generate these inputs, but 
also other sources such as model-based spatial optimisation approaches or even 
randomly generated masks could be used instead. Note that here, the optimised
mask values are fed into the surrogate solver instead of the original ones.

All three types of networks use a similar U-net structure~\cite{RFB15} that we 
discuss in more detail in Section~\ref{sec:unets}. In the following sections on 
the individual networks, we only discuss deviations from this standard U-net 
architecture.

Deploying our networks for practical applications comes down to first applying 
the mask network to the input image. The resulting mask is then optionally 
fed into the tonal network together with the original. This yields the 
complete known data for homogeneous diffusion inpainting. The surrogate solver 
is never used in an evaluation scenario. Instead, we use model-based inpainting.


\subsection{The Surrogate Inpainting Network}

To train our mask and tonal networks, we require backpropagation from
inpainting results. For instance, this could be achieved by translating
a classical discrete implementation of a diffusion process into
a neural network~\cite{ASAPW21}, which results in a sequence of ResNet
\cite{HZRS16} blocks. However, this might require very
deep networks to reach the steady state of the diffusion process since
the number of ResNet blocks is tied to the diffusion time in such a 
scenario. Instead, we propose an alternative that approximates 
inpainting results more efficiently by also having access to the ground
truth.

The \emph{surrogate inpainting network} $\mathcal{I}$ takes known data 
specified in terms of the locations in a binary or non-binary mask $\bm c$ 
and pixel values $\bm g$ as an input. 
Note that these known values do not necessarily need to coincide with the 
corresponding data in the original $\bm f$. In addition, it has access to the
full known image $\bm f$. This network will only be used during training, and 
for evaluation, a model-based solver is responsible for the inpainting. 
Therefore, having access to the unknown pixels in $\Omega \setminus K$ eases 
the networks task and does not compromise the validity of data optimisation in 
any way.

The reconstruction $\bm u = \mathcal{I}\left(\bm f, \bm g, \bm c\right)$ 
should solve the discrete inpainting equation
\begin{equation}
    \label{eq:inpdisc}
    \left(\bm I - \bm C\right) \bm A \bm u - \bm C \left(\bm u -\bm 
    g\right) = \bm 0,
\end{equation}
which is a discretised version of Eq.~\eqref{eq:inp}. The finite difference 
discretisation of the Laplacian is represented by the matrix $\bm A \in 
\mathbb{R}^{n_x n_y \times n_x n_y}$ and $\bm C \in [0,1]^{n_x n_y 
    \times n_x n_y}$ is a diagonal matrix containing the mask entries.

Since the network aims at simulating a numerical solver for 
Eq.~\eqref{eq:inpdisc}, we follow the ideas of Alt et al.~\cite{ASAPW21} and 
define a corresponding \emph{residual loss}
\begin{equation}\label{eq:resloss}
    \mathcal{L}_R \!\left(\bm u, \bm g, \bm c\right) = \frac{1}{n_xn_y} 
    \| \left(\bm I - \bm C\right) \bm A \bm u - \bm C \left(\bm u -\bm 
    g\right)\|_2^2 \,.
\end{equation}
Here $\|\cdot\|_2$ denotes the Euclidean norm.
Note that the inpainting network is explicitly \emph{not} trained to minimise 
any reconstruction loss w.r.t. the original $\bm f$. The residual loss only 
makes sure that the networks produces a good approximation of  homogeneous
diffusion inpainting given the mask $\bm c$ and the pixel values $\bm g$.  
It follows similar principles as deep energy approaches~\cite{GFE21}. 
This ensures that the surrogate solver's access to the full original 
image does not skew the data optimisation.


\subsection{The Mask Network}
\label{sec:masknet}

Given the original image $\bm f$, our \emph{mask network} $\mathcal{M}$ outputs 
positional data in terms of the mask $\bm c = \mathcal{M}(\bm 
f)$ with a density $d$. 

\subsubsection{Non-binary Mask Networks}
\label{sec:nonbinary}

In the approach from our conference paper \cite{APW22},
our network outputs non-binary masks with values in $[0,1]$.
Our goal is that the choice of $\bm c$ should yield the best possible 
inpainting. Therefore, our network is equipped with an \emph{inpainting loss} 
that measures the deviation of the reconstruction $\bm u$ from the original 
$\bm f$ in terms of
\begin{equation}\label{eq:inploss}
    \mathcal{L}_I \!\left(\bm u, \bm f\right) = 
    \frac{1}{n_x n_y} \|\bm u -\bm f\|_2^2 \,.
\end{equation}
While this loss establishes a 
connection between mask positions and reconstruction quality, it does not 
address the density. To this end, we apply a sigmoid activation at the last
layer of our mask U-net, which limits the non-binary mask outputs to $[0,1]$.
If the  preliminary mask $\hat{\bm c}$ exceeds the target density $d$, we 
rescale it according to
\begin{equation}
    \label{eq:rescale}
    \bm c = \frac{d  \hat{\bm c}}{\frac{\|\hat{\bm c} \|_1}{n_x n_y} + 
    \varepsilon} \, .
\end{equation}
With $\varepsilon=10^{-5}$ we avoid rounding issues for very low
estimated mask densities and potential division by zero.

During training, our network passes on the non-binary confidence values. 
Values close to $1$ indicate that the mask network sees this position
as highly important, and a value close to $0$ marks unimportant positions.
For practical applications, however, we still require binary masks. These
can be extracted with a simple postprocessing: Interpreting the confidence
values as a probability, we perform a weighted coin flip for each confidence
value.

Our experiments show that this non-binary mask optimisation creates a
challenging energy landscape. During the training process, the mask network
can get stuck in local minima that assign equal confidence to every mask
pixel. Combined with the coin flip, this can lead to a uniform random mask.
As a remedy, we propose an additional \emph{mask loss}  $\mathcal{L}_M$ that 
acts as a regulariser by penalising the inverse variance
\begin{equation}
    \label{eq:maskreg}
    \mathcal{L}_M \!\left(\bm c\right) = \alpha \left(\sigma^2_{\bm 
        c} + \varepsilon\right)^{-1}
\end{equation}
As in Eq.~\eqref{eq:rescale}, $\varepsilon$ avoids division by zero.
The regularisation parameter $\alpha$ balances the influence of the mask
loss with the inpainting loss.
Not only does this discourage flat masks with equal confidence in
every pixel, but it also encourages confidence values close to $0$ and $1$.
This yields the additional benefit of a closer approximation of 
binary masks during training.

\subsubsection{Binary Mask Networks}

Recently, strategies for deep data optimisation of neural network-based 
inpainting have been proposed that also allow direct output of binary 
masks~\cite{Pe22}. This constitutes a challenge since the binarisation of real 
input values is a non-differentiable operation. However, end-to-end approaches 
that also learn the inpainting benefit from this binarisation, since the 
training of the inpainting network tends to be biased by a non-binary mask 
input. This leads to worse results during deployment of the inpainting network. 

For our own strategy, we investigate two different alternatives for direct 
binarisation and evaluate their performance in Section~\ref{sec:ablation}. 

\medskip
\noindent
\textbf{Strategy 1: Quantisation.} First, we directly adopt
the strategy of Peter~\cite{Pe22}: We interpret binarisation of $x \in \mathbb 
R$ by hard rounding $x \mapsto \lfloor c + 0.5 \rfloor$ as very coarse 
quantisation. Theis et  al.~\cite{TSCH17} have shown that simply approximating 
the derivative by $1$ yields very good results among more sophisticated 
alternatives.

For this strategy, the variance-based regularisation from Eq.~\eqref{eq:maskreg}
is not necessary. However, the enforcement of the target density via rescaling 
from the non-binary approach also does not work in this case. Therefore, we 
define the mask loss directly as the deviation from the target density $d$ 
according to
\begin{equation}\label{eq:maskloss}
    \mathcal{L}_M \!\left(\bm c\right) = 
    \left\lvert\,\frac{\|\bm c\|_1}{n_x n_y}- d\,\right\rvert 
    \,.
\end{equation}
Since the mask contains only binary values, the 1-norm  $\|\cdot \|_1$ yields 
the number of mask points and thus the mask loss measures the deviation from 
the target density $d$. While the non-binary strategy does not require 
a density loss, we found in our experiments that it can have a stabilising 
effect on training if added to the regulariser loss from Eq.~\eqref{eq:maskreg}.

\medskip
\noindent
\textbf{Strategy 2: Coin Flip.} Instead of quantisation, we can also modify
our non-binary approach to output binary masks. We keep the regularisation
mask loss and rescaling from Section~\ref{sec:nonbinary}, yielding a non-binary
confidence mask. However, during training, we directly add the coin flip
binarisation. This can be seen as an alternative quantisation approach instead
of the rounding operation in Strategy 1. We apply the same synthetic gradient
as in the first binary mask approach.

\medskip
In Section~\ref{sec:ablation} we evaluate the binary and non-binary alternatives
for mask generation in an ablation study.


\subsection{The Tonal Network}

Finally, our tonal network takes both the original image $\bm f$ and a mask 
$\bm c$ as an input. The mask can either originate from the mask network or an 
external source. 

Fortunately, we do not require binarisation layers, since the input masks are
already binary. Furthermore, the mask density is already fixed. Therefore, the 
tonal network uses  the U-net described in 
Section~\ref{sec:unets} without further need for modifications. It  
feeds the optimised pixel values $\bm g = \mathcal{T}(\bm f, \bm c)$ into the 
inpainting loss from Eq.~\eqref{eq:inploss}.

The residual network is trained with the residual loss w.r.t. the 
optimised known data $\mathcal{L}_R \!\left(\bm u, \bm g, \bm c\right)$ as 
well. 
While this works well, we have found in our experiments that the 
training of the surrogate solver can be stabilised by also minimising
the residual $\mathcal{L}_R \!\left(\bm u, \bm f, \bm c\right)$ w.r.t. 
the original known data. This provides a fixed reference point for the
residual solver, since in contrast to $\bm g$, the known data from $\bm f$
is not influenced by the training progress of the tonal network. This
prevents the training of the residual solver from getting trapped in local 
minima.


\subsection{Network Architecture}
\label{sec:unets}

\begin{figure*}[t]
    \centering
    \includegraphics[width=\linewidth]{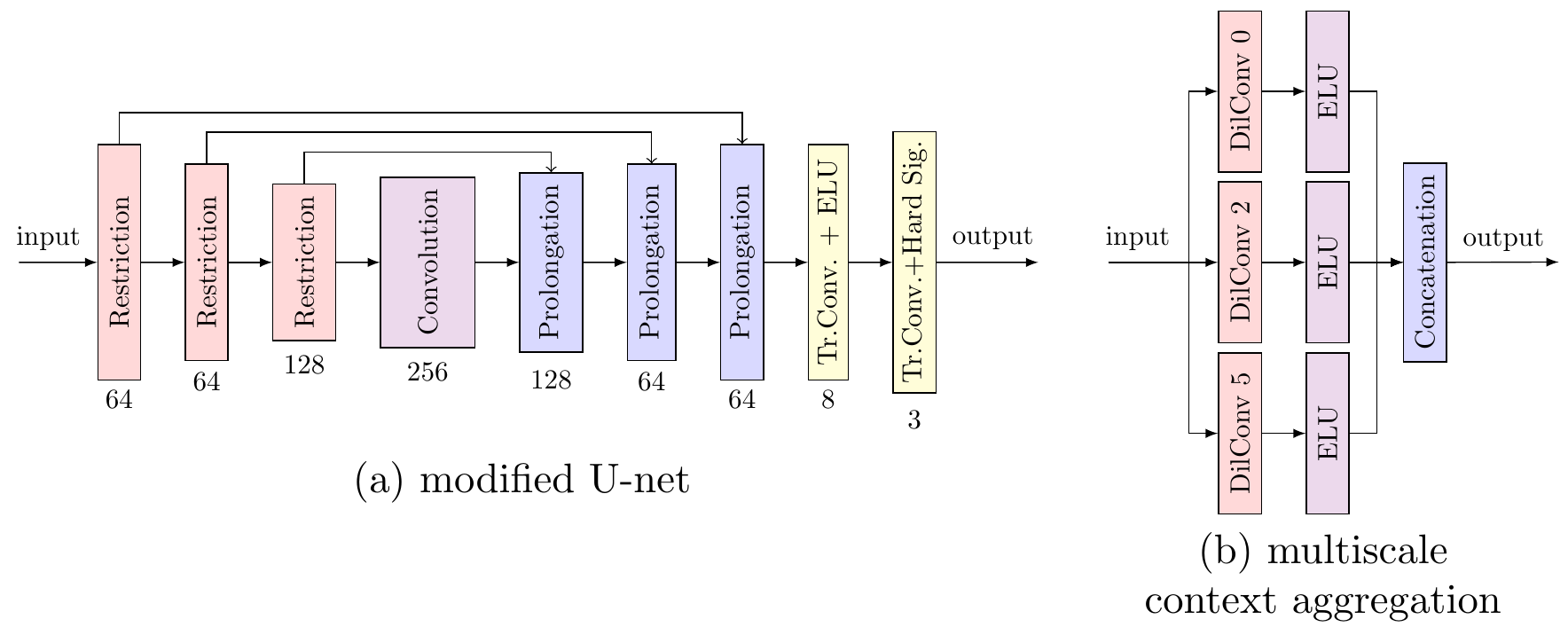}
    \caption{
        \textbf{Modified U-Net Architecture.} 
        (a) The original inputs are 
        subsampled by max pooling three times, pass through the bottleneck and 
        are then prolongated by upsampling again to the finest scale. Exchange 
        of information between scales is implemented by skip connections. Two 
        post processing blocks with transposed convolutions conclude the 
        pipeline. 
        The numbers below each context aggregation block indicate the number of 
        channels. 
        (b)  In a context aggregation block, three parallel dilated convolution 
        increase  the receptive field of the filter. All results are
        concatenated together. 
        \label{fig:unet}
    }
\end{figure*}

For all three networks, we use a U-net~\cite{RFB15} architecture, 
since U-nets implement the core principles of multigrid 
solvers for PDE-based inpainting~\cite{ASAPW21}. 
This makes them a perfect fit for the surrogate solver. U-nets and multigrid 
have in common that they operate on multiple scales, first restricting the 
image in multiple stages down to the coarsest scale and then prolongating it 
again to the finest scale. We follow this general structure in 
Fig.~\ref{fig:unet}(a). 

However, in contrast to our conference paper~\cite{APW22}, we also rely on 
modifications to the standard U-net approach that were first used for 
inpainting by Va\v{s}ata et al.~\cite{VHF21}. They replace traditional 
convolutional layers by multiple parallel dilated convolutions with dilation 
factors 0, 2, and 5 followed by ELU activations. As shown in 
Fig.~\ref{fig:unet}(b), the results are concatenated to a joint output. This 
so-called multiscale context aggregation was originally designed by Yu and 
Koltun~\cite{YK16} to increase the receptive field for segmentation. We 
discuss its benefits for our application in Section~\ref{sec:context} with an 
ablation study.

For restriction, we also use context aggregation~\cite{YK16} with $5\times5$ 
dilated convolutions followed by a $2 \times 2$ max pooling. The corresponding 
prolongation uses the same structure, but with $5\times5$ transposed 
convolutions and $2 \times 2$ upsampling. Two context aggregation blocks  
without any upsampling or max pooling perform postprocessing on the coarsest 
scale. The final hard sigmoid activation limits the results to the 
original image range $[0,1]$. Only in the case of our binary mask networks, 
this is followed by a quantisation or coin flip binarisation layer. 
As commonly the case in multiscale architectures, the number of channels
increases for coarser scales. It ranges from 64 to 256 (see 
Fig.~\ref{fig:unet}(a) for details), which is half of the channel bandwidth 
used by Va\v{s}ata et al.~\cite{VHF21}. In Section~\ref{sec:ablation} we 
have verified that such smaller networks suffice for our task. 


\section{Experimental Evaluation}\label{sec:experiments}

After an overview of the technical details of our evaluation in 
Section~\ref{sec:exp_setup}, we justify our design decisions for the networks
with an ablation study in Section~\ref{sec:ablation}. 
We compare with model-based approaches for spatial optimisation in 
Section~\ref{sec:exp_spatial} and with tonal optimisation methods in 
Section~\ref{sec:exp_tonal}. In both cases, we assess reconstruction quality 
and speed.


\subsection{Experimental Setup}
\label{sec:exp_setup}

Unless stated otherwise, all networks rely on the modified U-net architecture
from Section~\ref{sec:unets} with $\approx 2.9$ million parameters per network.

All of our networks have been trained on an Intel 
Xeon E5-2689 v4 CPU (2 cores), together with an Nvidia Pascal P100 16GB GPU.
For training, we use a subset of $100,000$ images randomly sampled from 
ImageNet~\cite{DDSL+09} by Dai et al.~\cite{DCPC19} and  the corresponding
validation dataset containing $1,000$ images. We use centre crops to 
reduce the size of the images, thus speeding up the training process. For model 
selection we crop to $64 \times 64$, while the remainder of the experiments are 
performed on size $128 \times 128$.
All networks were with the Adam optimiser~\cite{KB15} and a learning rate of $5 
\cdot 10^{-5}$. We used 50 epochs for the spatial experiments, and 100 for 
tonal experiments.
For evaluation, we used an AMD Ryzen 7 5800X CPU equipped with an Nvidia RTX 
3090 24GB GPU. We performed model selection based on the lowest achieved 
inpainting error on the validation set. Our test set is based on all 500 images 
of the BSDS500 database~\cite{AMFM11}. These were centre cropped to size 
$128 \times 128$ in order to fit the size of the training data. The cropping 
also speeds up the model-based competitors and thus allows us to compare 
with them on a larger variety of images. We measure qualitative results 
with the peak signal-to-noise ratio (PSNR).

We compare with three spatial optimisation methods. The analytic approach
by Belhachmi et al.~\cite{BBBW08} (AA) acts as a representative of very fast
spatial optimisation. It is implemented with Floyd-Steinberg 
dithering~\cite{FS76} of the Laplace magnitude. Probabilistic sparsification 
(PS) in combination with a non-local pixel exchange (NLPE) provides qualitative 
benchmarks. These methods have been implemented with a conjugate gradient 
solver, ensuring convergence up to a relative residual of $10^{-6}$ for the 
diffusion inpainting. NLPE is run for $5$ so-called cycles, each consisting of 
$\|\bm c \|_1$ iterations.


\subsection{Ablation Study}
\label{sec:ablation}

In the following, we first evaluate different architectures and design
principles, to select the best among those for the comparison
with model-based approaches.

\subsubsection{Network Architecture}
\label{sec:context}

Compared to the standard U-net architecture in our conference 
publication~\cite{APW22}, the modified U-net from Section~\ref{sec:unets} 
benefits from the context aggregation and more sophisticated postprocessing
layers after upsampling to the finest scale. 
In \cite{APW22}, we used sequential $3\times3$ convolutions on each scale. 
Therefore, propagation of information over larger distances works mainly via 
downsampling to coarse scales and upsampling. 
On each individual scale, the receptive field of the simple convolutions
is relatively small. In contrast, the context aggregation allows our network
to perceive larger regions of the image on each individual scale.
Our evaluation in Table~\ref{fig:ablation}(a) contrasts these 
modifications with the standard U-net using a similar total amount of weights. 
The modifications yield up to 2.3 dB improvement w.r.t. PSNR, especially on 
challenging very sparse masks.

We also evaluated other modifications to the U-net structure such as gated 
convolutions, but the context aggregation yielded the best combination of good 
qualitative performance and stability during training.

In Table~\ref{fig:ablation}(b), we compare the full size U-net proposed
by Va\v{s}ata et al.~\cite{VHF21} with our leaner version from 
Section~\ref{sec:unets} on $128\times128$ color images. The large U-net 
uses twice the amount of channels in relation to Fig.~\ref{fig:unet}(a) in all 
but the last two postprocessing layers. This results in $\approx 11.5$ million 
parameters, compared to our significantly lower  $\approx 2.9$ million. 
The larger network does not yield a qualitative advantage over a wide 
range of densities. The PSNR results of the large network only deviate 
marginally from those of the small masknet in Table~\ref{fig:ablation}(b).
However, it increases training times from 43 min to 93 min per 
epoch. Therefore, we use our lean nets instead.


\begin{table}[t]
    \centering
    \begin{minipage}[c]{0.6\textwidth}
        \centering
        \tabcolsep5pt
        \begin{tabular}{c|ccc}
            & \multicolumn{3}{c}{PSNR (dB)}\\
            density & 1\%  & 5\%  & 10\%  \\\hline
            non-binary \cite{APW22} & 19.40  & 25.45 & 28.34\\
            our non-binary & \textbf{21.72}  & \textbf{25.92} & \textbf{29.06}\\
            coinflip  & 18.61 & 24.99 & 22.58 \\ 
            binary  & 20.08 & 24.00 & 26.04 
        \end{tabular}
    \end{minipage}%
    \begin{minipage}[c]{0.4\textwidth}
        \centering
        \tabcolsep5pt
        \begin{tabular}{c|ccc}
            & \multicolumn{3}{c}{PSNR (dB)}\\
            density & 1\%  & 5\%  & 10\%  \\\hline
            small & 21.10  & 25.48 & 28.58\\
            large & 21.04 & 25.64 & 28.63
        \end{tabular}
    \end{minipage}

    \begin{minipage}[c]{0.6\textwidth}
        \centering
        \vspace{1mm}
        (a) binary vs. non-binary
    \end{minipage}%
    \begin{minipage}[c]{0.4\textwidth}
        \centering
        (b) small vs. large Masknet
    \end{minipage}\\[2mm]
    \caption{\textbf{Ablation Experiments.} All experiments have been 
    conducted on $64 \times 64$ grey value centre crops from 
    BSDS500~\cite{AMFM11} with mask densities $1\%$, $5\%$, and $10\%$. 
    (a) The non-binary masks outperform both binary options qualitatively 
    over the full range of mask densities. In addition, our modified 
    network architecture and training methodology outperforms the earlier 
    non-binary mask network~\cite{APW22}. The coinflip variant showed 
    instabilities during training for high densities and did thus not yield 
    satisfying results for 10\%. (b) Reducing the number of channels in the 
    modified U-net by a factor $2$ does not deteriorate the 
    quality.\label{fig:ablation}}
\end{table}


\subsubsection{Non-binary vs. Binary Masks}

In Section~\ref{sec:masknet} we have proposed three possible output options for 
our mask networks: non-binary masks, binary masks based on quantisation, and 
binary masks produced by a coin flip. For full deep learning based approaches
\cite{Pe22}, the binarisation during training is a key component of their 
architecture.

Surprisingly, our ablation study in Table~\ref{fig:ablation}(a) paints a 
different picture: The non-binary mask network clearly outperforms both binary 
options. This results from a key difference in our method compared to full deep
learning approaches. Using a non-binary mask while simultaneously training an 
inpainting network introduces a bias. This deteriorates inpainting quality 
during testing \cite{DCPC19}. However, our surrogate solver is only deployed 
during training and is not coupled directly to an inpainting loss. It merely
approximates diffusion-based inpainting. During testing, we use a model-based 
implementation of homogeneous diffusion inpainting.

Therefore, we benefit from a non-binary mask network that does not rely on 
synthetic gradients for binarisation layers. Consequentially, we use the 
non-binary variant for our comparisons with model-based data optimisation.


\begin{figure}[t]
\tabcolsep1pt
\centering
\includegraphics[width=0.23\linewidth]{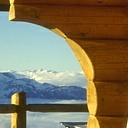} 
\\[-0.5mm]
original image 61034\\ BSDS500~\cite{AMFM11}\\[3mm]
\begin{tabular}{ccccc}
& AA~\cite{BBBW08} & PS~\cite{MHWT12} & PS+NLPE~\cite{MHWT12} & our 
Masknet \\
\rotatebox{90}{\hspace{0.8cm} mask} & 
\includegraphics[width=0.23\linewidth]{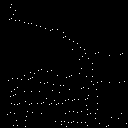}
 &
\includegraphics[width=0.23\linewidth]{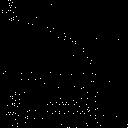}
 &
\includegraphics[width=0.23\linewidth]
{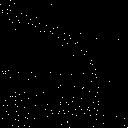} &
\includegraphics[width=0.23\linewidth]{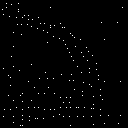}
  \\
\rotatebox{90}{\hspace{0.4cm} spatial only} & 
\includegraphics[width=0.23\linewidth]{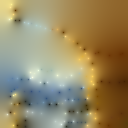}
 &
\includegraphics[width=0.23\linewidth]{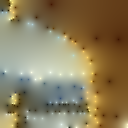}
 &
\includegraphics[width=0.23\linewidth]{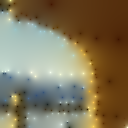}
 &
\includegraphics[width=0.23\linewidth]{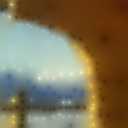}
   
\\[-0.5mm]
& PSNR: 13.73 & PSNR: 16.96 & PSNR: 19.90 & PSNR: 19.99 \\[1mm]
\rotatebox{90}{\hspace{0.1cm} spatial + tonal} &  
\includegraphics[width=0.23\linewidth]{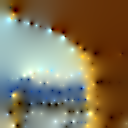}
 &
\includegraphics[width=0.23\linewidth]{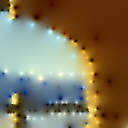}
 &
\includegraphics[width=0.23\linewidth]{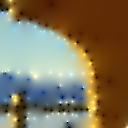}
 &
\includegraphics[width=0.23\linewidth]
{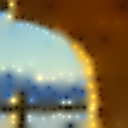}  \\[-0.5mm]
& PSNR: 18.55 & PSNR: 21.35 & PSNR: 22.27 & PSNR: 21.98 \\
\end{tabular}\\[1mm]
\caption{\textbf{Visual Comparison for 1\% Mask Density.} Visually, 
probabilistic sparsification (PS) in combination with non-local pixel exchange
(NLPE)~\cite{MHWT12}, and our network-based approach significantly outperform
the analytic approach (AA)~\cite{BBBW08}. For such sparse masks, the visual 
impact of tonal optimisation is apparent.
\label{fig:visual_01}}
\end{figure}


\begin{figure}[t]
\centering
\includegraphics[width=0.23\linewidth]{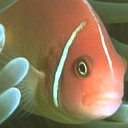} 
\\[-0.5mm]
original image 210088\\ BSDS500~\cite{AMFM11}\\[3mm]
\tabcolsep1pt
\begin{tabular}{ccccc}
& AA~\cite{BBBW08} & PS~\cite{MHWT12} & PS+NLPE~\cite{MHWT12} & our 
Masknet\\
\rotatebox{90}{\hspace{0.8cm} mask} & 
\includegraphics[width=0.23\linewidth]{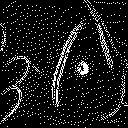}
 &
\includegraphics[width=0.23\linewidth]{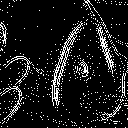}
 &
\includegraphics[width=0.23\linewidth]
{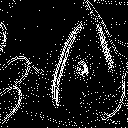} &
\includegraphics[width=0.23\linewidth]{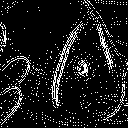}
  \\
\rotatebox{90}{\hspace{0.4cm} spatial only} & 
\includegraphics[width=0.23\linewidth]{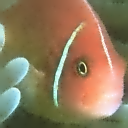}
 &
\includegraphics[width=0.23\linewidth]{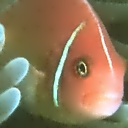}
 &
\includegraphics[width=0.23\linewidth]{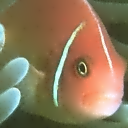}
 &
\includegraphics[width=0.23\linewidth]{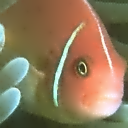}
   
\\[-0.5mm]
& PSNR: 29.25 & PSNR: 32.27 & PSNR: 33.72 & PSNR: 32.83 \\[1mm]
\rotatebox{90}{\hspace{0.1cm} spatial + tonal} &  
\includegraphics[width=0.23\linewidth]{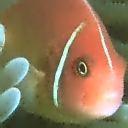}
 &
\includegraphics[width=0.23\linewidth]{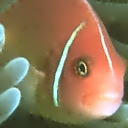}
 &
\includegraphics[width=0.23\linewidth]{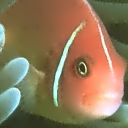}
 &
\includegraphics[width=0.23\linewidth]
{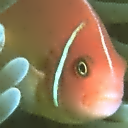}  \\[-0.5mm]
& PSNR: 32.67 & PSNR: 34.38 & PSNR: 35.55 & PSNR: 34.17  \\
\end{tabular}\\[1mm]
\caption{\textbf{Visual Comparison for 10\% Mask Density.} 
Also at high density, our mask network yields results that are comparable to 
the probabilistic approaches PS and PS+NLPE~\cite{MHWT12}. The visual gap 
towards the analytic approach (AA)~\cite{BBBW08} is smaller, but still 
noticable.\label{fig:visual}}
\end{figure}


\subsection{Spatial Optimisation}
\label{sec:exp_spatial}


\begin{figure}[t]
    \centering
    \begin{minipage}[c]{0.5\textwidth}
        \centering
        \includegraphics[width=\linewidth]{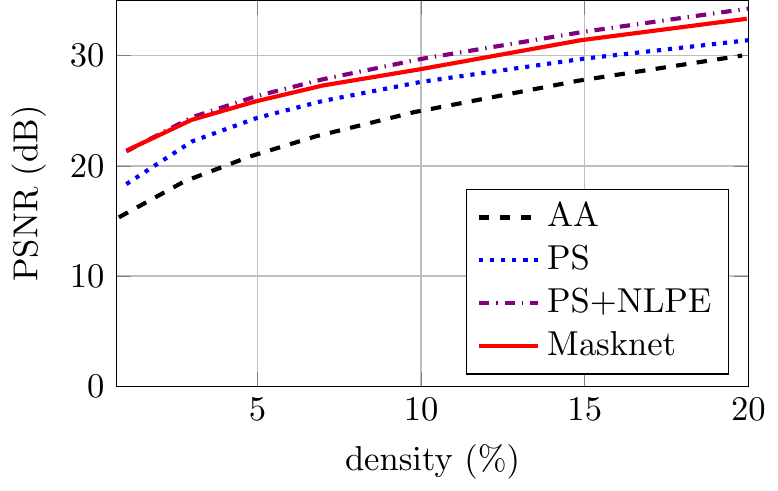}
    \end{minipage}%
    \begin{minipage}[c]{0.5\textwidth}
        \centering
        \includegraphics[width=\linewidth]{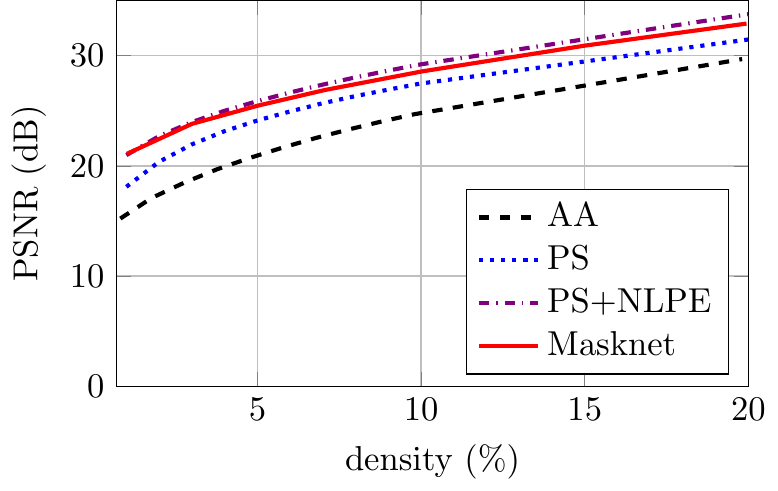}
    \end{minipage}
    
    \begin{minipage}[c]{0.6\textwidth}
        \centering
        (a) BSDS500 grey
    \end{minipage}%
    \begin{minipage}[c]{0.4\textwidth}
        \centering
        (b) BSDS500 colour
    \end{minipage}\\[2mm]
\caption{\textbf{Spatial Optimisation.} (a) On grey level images, our 
network consistently outperforms the analytic spatial optimisation (AA) 
\cite{BBBW08} and probabilistic sparsification (PS)~\cite{MHWT12}. Especially 
for lower densities, Masknet results rival the quality of PS with non-local 
pixel exchange (NLPE)~\cite{MHWT12} as postprocessing.
(b) For colour images, our mask network also closely approximates the quality 
of PS+NLPE for the whole range from 1\% to 20\%. \label{fig:spatial}}
\end{figure}


In our conference paper~\cite{APW22}, we have shown that our approach
yields similar results as probabilistic sparsification~\cite{MHWT12}
on a small dataset of five greyscale images. Here we extend the evaluation of 
our improved networks to the significantly larger greyscale BSDS500 database in 
Fig.~\ref{fig:spatial}(a) and double the range of evaluated mask densities to 
20\%. Our mask network not only consistently outperforms both the analytic 
approach~\cite{BBBW08} (AA) and probabilistic sparsification~\cite{MHWT12} 
(PS), but very closely approximates the quality of PS+NLPE.

The same ranking also applies in the case of the full colour version of BSDS500 
in Fig.~\ref{fig:spatial}(b). Thus, our mask network rivals the best 
model-based approach in the comparison. Visually, it yields similar results as 
the probabilistic methods in Fig.~\ref{fig:visual_01} and 
Fig.~\ref{fig:visual}. Especially for low densities, there is a large quality 
gap between the analytical approach and all other competitors.

\begin{figure*}[t]
\centering
\tabcolsep 2pt
\begin{tabular}{cc}
    \includegraphics[width=0.51\linewidth]{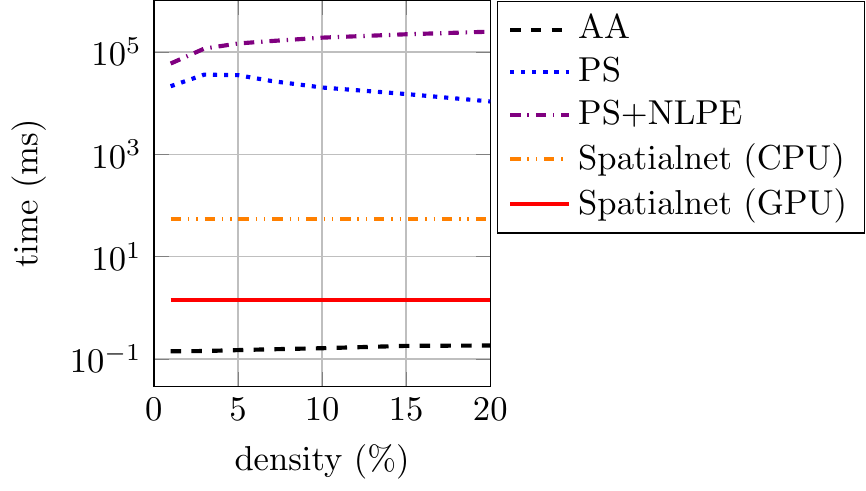} &
    \includegraphics[width=0.48\linewidth]{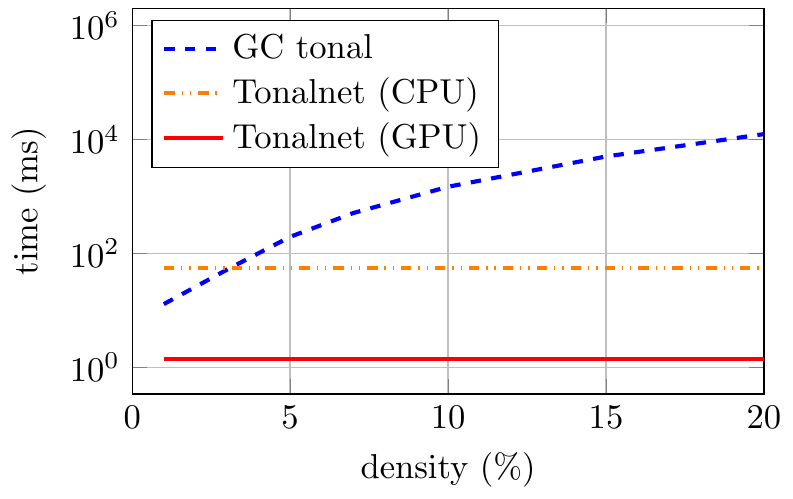} \\
    (a) runtime spatial & (b) runtime tonal
\end{tabular}\\[2mm]

\caption{
    \textbf{Runtime Comparison with Logarithmic Time Axis.} 
    (a) The analytic approach (AA) \cite{BBBW08} is the fastest method, 
    followed closely by our Masknet on the GPU and CPU. These three methods all 
    have a constant speed independently of mask density. The speed of PS and 
    PS+NLPE~\cite{MHWT12} is density dependent. Overall, our methods are 
    consistently faster by several orders of magnitude compared to 
    probabilistic approaches.
    (b) The situation for tonal optimisation is comparable. The Green's 
    function-based solver~\cite{Ho17} becomes increasingly slower with 
    rising mask density. Only for very low densities its speed is comparable to 
    our tonal network on the CPU. The networks have constant runtime 
    independent of density and are faster by 1 to 5 orders of magnitude. 
    \label{fig:runtime}
}
\end{figure*}

Even though our mask net offers a similar quality as PS+NLPE, it requires
significantly less computational time since it does not rely on any inpaintings 
during inference. On the CPU, it accelerates mask computation by up to a factor 
3500 and even up to a factor 140.000 on the GPU in Fig.~\ref{fig:runtime}(a).
Only the analytic approach is faster with $\approx 0.2$ ms. However, there 
the speed comes at the cost of a significantly diminished quality. Without 
compromising on quality,  our mask net is also real-time capable with $1.4$ ms 
on GPU and $55$ ms on CPU.

Thus, our mask network reaches our goal of providing an easy to use, 
parameter-free spatial optimisation which approximates the quality
of stochastic methods at a computational cost close to the instantaneous 
analytic approach. 


\begin{figure}[t]
\centering
\tabcolsep0.5pt
\begin{tabular}{cc}
    \includegraphics[width=0.5\linewidth]{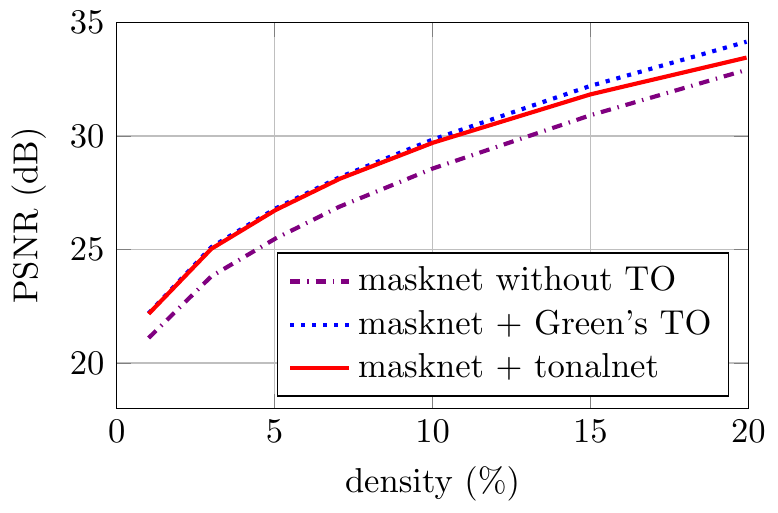}
    &
    \includegraphics[width=0.5\linewidth]{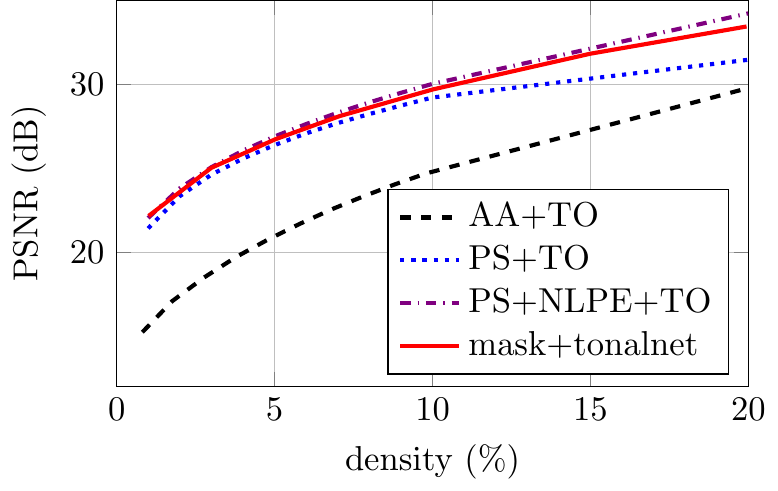} \\
    (a) neural vs. model-based  & (b) full optimisation comparison
\end{tabular}\\[2mm]
\caption{
    \textbf{Tonal Optimisation.} (a) We compare our tonal network with 
    the Green's function approach~\cite{Ho17} on masks from our mask network. 
    Our tonal optimisation (TO) reaches a comparable quality up to 15\% known 
    data.
    (b) In a comparison that combines the spatially optimised mask with 
    tonal optimisation, our full network approach yields competitive results to 
    PS+NLPE combined with tonal optimisation. All model-based approaches use 
    the Green's function approach~\cite{Ho17} for tonal optimisation.
    \label{fig:tonal}
}
\end{figure}


\subsection{Tonal Optimisation}
\label{sec:exp_tonal}

In Fig.~\ref{fig:tonal}(a) we compare our tonal network with the Green's 
function approach of Hoffmann~\cite{Ho17} on the masks obtained from our mask 
network. Especially for sparse known data, our deep tonal optimisation reaches 
a similar quality as the model-based approach. Only above 15\%, the 
improvements over the unoptimised data from the mask net decline.

Our results in Fig.~\ref{fig:visual_01} show that our network approach also 
remains competitive to PS+NLPE when adding tonal optimisation. Here we apply 
the tonal network for our own deep learning method and the Green's function 
optimiser for all model-based competitors.

As for spatial optimisation, our tonal network offers a viable alternative for 
time critical applications. Fig.~\ref{fig:runtime}(b) shows that the 
computational cost of the Green's function approach grows significantly with 
the number of mask values that need to be optimised. In contrast, the 
computational time of the tonal network is independent of the mask density. For 
densities larger than 5\%, speed-ups by multiple orders of magnitude can be 
achieved with our mask net.

Thus, a combination of our spatial and tonal networks is a viable option for 
real-time applications that does not require to sacrifice quality for speed. 

\section{Conclusions}\label{sec:conclusion}
Our data optimisation approach merges classical inpainting with partial 
differential equations and deep learning with a surrogate solver. 
This allows us to select both position and values of known data for 
homogeneous diffusion inpainting that minimise the reconstruction error. 

With this new strategy for sparse data  optimisation we obtain real-time
results in hitherto unprecedented quality. They yield reconstructions that
rival the results of probabilistic sparsification with postprocessing
by non-local pixel exchange and tonal optimisation. Simultaneously,
they are reaching the near instantaneous speed of the qualitatively inferior 
analytic approach. This improvement of computational time by multiple orders of 
magnitude at comparable quality demonstrates the high potential of a fusion 
between model- and learning-based principles. We see this as a milestone on our 
way to bring the best of both worlds together in the area of inpainting and 
data optimisation.

In the future, we plan to incorporate our framework into image compression 
codecs. Time-consuming spatial and tonal optimisation still present a 
bottleneck in this area. This holds true especially for practical applications 
with high demand for computational efficiency, such as video coding. While 
real-time decoding is already possible with diffusion 
\cite{KSFR07,PSMM15,APMW21}, the data selection during encoding will benefit 
from our deep optimisation.

\backmatter

\bmhead{Acknowledgments} This work has received funding from the European 
Research Council (ERC) under the European Union's Horizon 2020 research and 
innovation programme (grant agreement no. 741215, ERC Advanced Grant 
IN\-CO\-VID).  

\bibliography{myrefs}

\end{document}